\documentclass[twocolumn,showpacs,preprintnumbers,amsmath,amssymb]{revtex4}
\usepackage{graphicx}
\usepackage{dcolumn}
\usepackage{bm}

\newcommand{\be}{\begin{equation}}
\newcommand{\ee}{\end{equation}}
\newcommand{\ba}{\begin{eqnarray}}
\newcommand{\ea}{\end{eqnarray}}
\newcommand{\bse}{\begin{subequations}}
\newcommand{\ese}{\end{subequations}}
\newcommand{\M}{{\cal {M}}}
\newcommand{\B}{{\cal {B}}}

\newcommand{\HH}{{\cal {H}}}

\newcommand{\CV}{{\cal {V}}}

\begin{document}

\title[Local...collapse of a self-gravitating magnetized Fermi gas.]
{ Local dynamics and gravitational collapse of a self-gravitating magnetized Fermi gas.}

\author{A. Ulacia Rey$^\dagger$, A. P\'erez Mart\'inez$^\dagger$ and Roberto A. Sussman$^\ddagger$}

\address{$^\dagger$Instituto de Cibern\'etica Matem\'atica y F\'{\i}sica
(ICIMAF). Calle E No-309 Vedado, cp-10400. Ciudad de La Habana,
Cuba.\\$^\ddagger$ Instituto de F\'{\i}sica, Universidad de
Guanajuato, Loma del Bosque 103, Leon, Guanajuato, 37150, Mexico (on
sabbatic leave from Instituto de Ciencias Nucleares, UNAM, Mexico D.F., 04510)}

\email{alain@icmf.inf.cu;aurora@icmf.inf.cu;sussman@nucleares.unam.mx}

\begin{abstract}
We use the Bianchi-I spacetime to study the local dynamics of a
magnetized self--gravitating Fermi gas. The set of Einstein-Maxwell
field equations for this gas becomes a dynamical system in a
4-dimensional phase space. We consider a qualitative study and
examine numeric solutions for the degenerate zero temperature case.
All dynamic quantities exhibit similar qualitative behavior in the
3-dimensional sections of the phase space, with all trajectories
reaching a stable attractor whenever the initial expansion scalar
$H_0$ is negative. If $H_{0}$ is positive the trajectories end up
in a curvature singularity that can be, depending on initial conditions,
 isotropic or anisotropic. In particular, if the initial magnetic field
 intensity is sufficiently large the collapsing singularity will always
 be anisotropic and pointing in the same direction of the field.
\end{abstract}
\maketitle
\section{Introduction}
White dwarfs and neutrons stars, as very dense objects found in
great abundance, are astrophysical laboratories to test physical
theories under intense gravity. In particular, we can consider as a
theoretical model the possibility of Magnetic White Dwarfs that
could be  endowed with extremely large magnetic fields, which, in
principle, can be stronger than those measured in earthbound
laboratories. It is very interesting to study the behavior of the
magnetic field lines, in its interplay with strong gravity, as such
a magnetic star undergoes gravitational collapse. Various
qualitative arguments show that under gravitational collapse the
field lines squeeze together making the field  stronger, even
resisting collapse.

Several models describing magnetic white dwarfs have been
constructed (see \cite{Lattimer1,Lattimer2,Kubica}) using Landau's
well known  argument setting appropriate mass limits for these
objects. Ideally, a consistent model of a magnetic white dwarf would
require considering numeric hydrodynamical modeling of
Einstein--Maxwell equations with (at the very least) axially
symmetric configurations that would have to comply with the
appropriate boundary conditions and equations of state of a
degenerate magnetized Fermi gas. However, we can still obtain
important information on the local dynamics of the self--gravitating
magnetized Fermi gas by considering the evolution of such a source
in a much more simplified spacetime geometry.

The type of equation of state that we consider for a magnetized
white dwarf is similar to that discussed in previous papers dealing
with a magnetized electron gas~\cite{Canuto1,Canuto2,Canuto3} (1968)
and later developed for more general
sources~\cite{Aurora1,Aurora2,Aurora3,Chakra1,Chakra2}(2000). These
articles discuss equations of state for strongly magnetized gases
endowed with anisotropic pressure.  Previous work in a Newtonian
framework noted the possibility in which the  pressure
$p_{\parallel}$ parallel to the magnetic field overtakes the
pressure $p_{\perp}$ perpendicular  to the field, suggesting  a
magnetic collapse in this direction. This brings us to think that a
general relativistic treatment of magnetized configurations will
yield a singularity associated to the collapse having an extended,
anisotropic, form along the direction of the magnetic field. Such
type of singularities are denoted as ``line'' or ``cigar''
singularities, as opposed to isotropic ''point--like''
singularities.

In order to investigate the local dynamics and collapse of a
magnetized Fermi gas under General Relativity, we consider one of
the simplest geometric configurations compatible with the
anisotropic pressure that characterizes these sources: the Bianchi I
model described in terms of a Kasner metric. The present paper
generalizes previous work along these lines~\cite{AAS}.

The main justification of dealing with the simplified Bianchi I
geometry is that the latter could provide a rough description of the
local dynamics of a fluid element of the magnetized gas in the
realistic configuration. If we consider local fluid elements far
from the boundary of the configuration, so that this volume
exchanges particles and energy with the rest of the system (seen as
a reservoir), then we can consider it like a local volume of a great
canonical distribution associated with the whole gas. While it is
evident that a lot of valuable information is lost by making such
simplifications, we can still get a rough description of the effects
of strong gravity on local physics.

Once we consider the Bianchi I spacetime with the Kasner metric as
the metric field associated with the magnetized gas, the
Einstein--Maxwell system of field equations reduce to an autonomous
system of four ordinary differential equations, leading to a
4--dimensional phase space that can be studied from a  qualitative
and numerical point of view using standard dynamical systems
techniques. The physical and geometric dimensionless variables of
this phase space are the single independent component of the shear
tensor $S$, the expansion scalar $\HH$, the normalized dimensionless
magnetic field and the chemical potential, $\beta$ and $\mu$. All
other quantities can be expressed in terms of these four basic
quantities. The numerical examination of the system sheds light on
the type of collapse singularities and their relation to specific
initial conditions.

The equation of state that we are using is strictly valid for
densities of the order of (at least)  $10^{7}\textrm{g cm}^{-3}$ and
up to $10^{15}\textrm{g cm}^{-3}$ expected in compact objects from
white dwarves to neutron stars. A gas of strongly magnetized and
highly degenerate fermions in a neutron star is in a state that
closely resembles superfluidity with near infinite conductivity (see
page 291 \cite{Shapiro1}). In these conditions the role of viscosity
is minor, though one can still consider the possibility of
dissipative or transport phenomena, such as dissipation of
rotational energy in electromagnetic and gravitational waves (see
\cite{Shapiro1,Shapiro2,Shapiro3}).

However, even if viscosity is not significant (at least for neutron stars), the main reason why
we are neglecting it (and other dissipative effects) is to keep a
mathematically tractable problem. We feel that treating the case of
thermal equilibrium is sufficient for a first approach, leaving the
study of dissipative transport phenomena for a future work.

The paper is organized as follows. In section II we present and
discuss the set of Einstein--Maxwell equations, the source of
anisotropy and the most appropriate form of the equations of state
for the magnetized Fermi gas. The qualitative dynamical analysis is
carried on in Section-III, defining a set of dimensionless
normalized variables, leading to a self--consistent autonomous
system of four ordinary differential equations. The numerical and
qualitative analysis of this system and the classification of the
types of collapse singularities are dealt with in Section-IV, while
the conclusion is given in section V. The main result from the
numerical analysis is that once we allow for a relativistic strong
gravity treatment it is always possible, for sufficiently large
magnetic field initial intensity, to obtain  the anisotropic
``cigar'' type of collapse singularity as hypothesized in previous
work~\cite{Aurora3} carried on along an intuitive Newtonian
framework work.

\section{Kasner Metric with Anisotropic Pressure}

The Kasner metric is among the simplest metrics compatible with the anisotropic
pressure associated with a magnetized source. This metric is given by:
\begin{equation}
{ds^2} \ = \
-c^2\,dt^2+A^2(t)\,dx^{2}+B^2(t)\,dy^2+C^2(t)\,d{z}^2.\label{KMWAP1}
\end{equation}
It is associated with a ``non--tilted'' Bianchi-I space time
\cite{DSIC}. For a comoving 4-velocity $u^{a}=\delta^{a}_{t}$,
where $u^{a}u_{a}=-1$, the 4--acceleration vanishes and the
expansion scalar $\Theta$ and the shear tensor $\sigma^{a}_{b}$ take
the forms:
\bse\label{KMWAP2}\ba
\sigma^{a}_{b} &=& {\textrm {diag}}[\sigma^{x}_{x},\sigma^{y}_{y},\sigma^{z}_{z},0],\\
 \Theta &=& \frac{\dot{A}}{A}+\frac{\dot{B}}{B}+\frac{\dot{C}}{C},
\ea\ese
where
\begin{subequations}
\begin{eqnarray}
\sigma^{x}_{x}&=&\frac{2\dot{A}}{3A}-\frac{\dot{B}}{3B}-\frac{\dot{C}}{3C},\
\
\sigma^{y}_{y}=\frac{2\dot{B}}{3B}-\frac{\dot{A}}{3A}-\frac{\dot{C}}{3C},\
\\\
\sigma^{z}_{z}&=&\frac{2\dot{C}}{3C}-\frac{\dot{A}}{3A}-\frac{\dot{B}}{3B},\
\ \ \ \ ( \sigma^{a}_{a}=0 ).
\end{eqnarray}\label{KMWAP3}
\end{subequations}
We consider as the source for this metric the following
stress-energy tensor:
\begin{equation}
T^{a}_{b}=(U+P)u^{a}u_{b}+P\delta^{a}_{b}+\Pi^{a}_{b}, \ \
P=p-\frac{2\B\M}{3}. \label{KMWAP4}
\end{equation}
where $\Pi^{a}_{b}$ is the anisotropic pressures tensor, $U$ the
energy density, $P$ the pressure, $\B$ the magnetic field and $\M$
the magnetization, all them are functions of the time. Notice that
the anisotropy is produced by the magnetic field $\B$. If this field
vanishes, {\textit ie}: $\B=0$, the stress-tensor reduces to that of
a perfect fluid tensor with isotropic pressure $P = p$. In the
general case $\B\ne 0$ the tensor $\Pi^{a}_{b}$ has the form
\begin{equation}
\Pi^{a}_{b}={\textrm {diag}}[\Pi,\Pi,-2\Pi,0], \ \ \ \Pi=-\frac{\B\M}{3},\ \ \
\Pi^{a}_{a}=0. \label{KMWAP5}
\end{equation} The Einstein field equations (EFE) associated with the Kasner metric
 and the stress-energy tensor are
 \begin{subequations}
\begin{eqnarray}
-G^{x}_{x}&=&\frac{\dot{B}\dot{C}}{BC}+\frac{\ddot{B}}{B}+\frac{\ddot{C}}{C}=-\kappa(p-\B\M),
\label{KMWAP6a}\\
-G^{y}_{y}&=&\frac{\dot{A}\dot{C}}{AC}+\frac{\ddot{A}}{A}+\frac{\ddot{C}}{C}=-\kappa(p-\B\M),
\label{KMWAP6b}\\
-G^{z}_{z}&=&\frac{\dot{A}\dot{B}}{AB}+\frac{\ddot{A}}{A}+\frac{\ddot{B}}{B}=-\kappa
p,\label{KMWAP6c}
\\
-G^{t}_{t}&=&\frac{\dot{A}\dot{B}}{AB}+\frac{\dot{A}\dot{C}}{AC}+\frac{\dot{B}\dot{C}}{BC}=\kappa
U.\label{KMWAP6d}
\end{eqnarray}
\label{KMWAP6}
\end{subequations}
where a dot denotes the derivative with respect to the proper time of
fundamental observers and $\kappa=8\pi G/c^4$. From the balance
equation $T^{ab}_{\ \ ;b}=0$ and the Maxwell equations $F^{ab}_{\ \
;b}=0$ and $F_{[ab;c]}=0$, we further have
\begin{subequations}
\begin{eqnarray}
\dot{U}&+&(p+U)\Theta-\B\M(\frac{\dot{A}}{A}+\frac{\dot{B}}{B})=0,
\label{KMWAP7a}\\
\frac{\dot{A}}{A}&+&\frac{\dot{B}}{B}+\frac{1}{2}\frac{\dot{\B}}{\B}=0.
\label{KMWAP7b}
\end{eqnarray}
\label{KMWAP7}
\end{subequations}
Since we need to construct a self-consistent system of first order
differential equations that can be solved numerically, it is
convenient to eliminate first and second derivatives of the metric
functions in the Einstein--Maxwell equations in terms of the
expansion scalar and the components of the shear tensor. Proceeding
along these lines, we combine equations
(\ref{KMWAP2}),(\ref{KMWAP3}), (\ref{KMWAP6}) and (\ref{KMWAP7}) to
eliminate the functions $A,B,C$ and their derivatives
$\dot{A},\ddot{A},\dot{B},\ddot{B},\dot{C},\ddot{C}$. After some
algebraic manipulation we arrive to the following constraint
\begin{equation}-(\Sigma^{y})^{2}=(\Sigma^{z})^{2}-\Sigma^{y}\Sigma^{z}+\frac{\Theta^{2}}{3}=\kappa U,\label{constr1}\end{equation}
plus the following set of 5 differential equations:
\begin{subequations}
\begin{eqnarray}
\dot{U}&+&(U+p)\Theta-\B\M(\frac{2}{3}\Theta-\Sigma^{z})=0,
\\
\dot{\Sigma^{y}}&=&-\frac{\kappa}{3} \B\M-\Sigma^{y}\Theta,
\\
\dot{\Sigma^{z}}&=&\frac{2\kappa}{3} \B\M-\Sigma^{z}\Theta,
\\
\dot{\Theta}&=&\kappa (\B\M-\frac{3}{2}p)-\frac{
\Theta^{2}}{2}-\frac{3}{2}((\Sigma^{y}+\Sigma^{z})^{2}-\Sigma^{y}\Sigma^{z}),
\nonumber \\ \\
\dot{\beta}&=&\frac{2}{3}\beta (3\Sigma^{z}-2\Theta),
\quad \textrm{with}\quad \beta\equiv \B/\B_{c}.
\end{eqnarray}
\label{KMWAP8}
\end{subequations}
where $\Sigma^{z}=\sigma^{z}_{z}$ is the independent component of
the shear tensor. While the shear tensor can be fully determined by
this single quantity, it will be convenient for our numerical
calculations further ahead to use two components of this tensor.

The functions $U, p$ and $\M$ are now given by the
equation of state for the gas;
 \be
 p= \lambda\,\, \Gamma_{p}(\beta,\mu), \ \ \ \B\M \ = \ \lambda\,
\beta\Gamma_{{\M}}(\beta,\mu), \ \ \ U \ = \ \lambda\,
\Gamma_{_U}(\beta,\mu). \label{KMWAP9} \ee
The $\Gamma $ functions depend of the parameter $\beta$, which is the magnetic
field normalized by $\B_ {c}$, and by the chemical potential, $\mu$,
normalized with the rest energy, both dimensionless quantities. The
constants $B_{c}$ and $\lambda$ are:
\begin{equation}
\B_{c}=\frac{m^{2}c^{3}}{e\hbar}, \ \ \ \ \
\lambda=\frac{mc^2}{4\pi^{2} \lambda^{3}_{c}}. \label{KMWAP10}
\end{equation}
 while $\lambda_{c}=\hbar/mc$ is the Compton wavelength. If we
consider an electrons gas, then $m=m_{e}$ and
$\lambda=3.86.10^{-11}{\textrm {cm}}$. Nevertheless, $B_{c}=4.414
\times 10^{13}$ G (for electrons too) which is a well known critical
value of a magnetic field. In all neutron or white dwarf stars older
than few second after formation, one can neglect the thermal
contributions to the pressure and energy density; thus we can set
$p(\beta,\mu,T)=p(\beta,\mu)$, the same for $U$ and $\M$. Typical
white dwarf temperatures satisfy $kT << E_{\textrm{Fermi}}$, where
$E_{\textrm{Fermi}}$ is the Fermi kinetic energy. So, the thermal
disorder $k\,T$ is not responsible for the pressure, the energy
density nor the magnetization in (\ref{KMWAP9}). It is now
convenient to introduce the form of the $\Gamma(\beta,\mu)$
functions for the degenerate case $(T=0)$, these are \cite{Aurora1}:
\begin{widetext}
\begin{subequations}
\begin{eqnarray}
\Gamma_{p}&=&\frac{a_{_0}}{3}\left(\mu^2-\frac{5}{2}\right)+\frac{1}{2}\textrm{arcsinh}\left(\frac{a_{_0}}{\mu}\right)+\beta\sum^{s}_{n=0}\alpha_{n}(a_{n}-b_{n}-c_{n}),\ \label{KMWAP11a} \\
\Gamma_{\M}&=&\sum^{s}_{n=0}\alpha_{n}(a_{n}-b_{n}-2c_{n}), \label{KMWAP11b} \ \ \ \ \\
\Gamma_{U}&=&a_{_0}\left(\mu^2-\frac{1}{2}\right)-\frac{1}{2}\textrm{arcsinh}\left(\frac{a_{_0}}{\mu}\right)+\beta\sum^{s}_{n=0}\alpha_{n}(a_{n}+b_{n}+c_{n}),
\qquad \textrm{where}\ \ \ \alpha_{n}=2-\delta_{0n}, \ \ n=0,1,..    \label{KMWAP11c} \\
\textrm{while} \ \ \ a_{n}&=&{\mu\sqrt{\mu^2-1-2n\beta}}, \qquad
b_{n}=\ln\left[\frac{(\mu+a_{n}/\mu)}{\sqrt{1+2n\beta}}\right],\qquad
c_{n}=2n\beta b_{n}, \qquad
s=I\left[\frac{\mu^2-1}{2\beta}\right].\label{KMWAP11c}
\end{eqnarray} \label{KMWAP11}\end{subequations}\end{widetext}
where $I[X]$ denotes the integer part of its argument $X$.

\section{Dimensionless Variables}

Consider now the following variables:
\begin{equation}
H=\frac{\Theta}{3}, \ \ \
\frac{d}{d{\tau}}=\frac{1}{H_{0}}\frac{d}{c\ dt}. \ \label{DV1} \ee
and the dimensionless functions: \be S^{y}=\frac{\Sigma^{y}}{H_{0}},
\ \ \ \ S^{z}=\frac{\Sigma^{z}}{H_{0}}, \ \ \ \
\Omega=\frac{\kappa\lambda\beta}{3H^{2}_{0}}, \ \ \ \
\HH=\frac{H}{H_{0}}. \label{DV2}
\end{equation}
where $S^{y}$ and $S^{z}$ are related to the $yy$ and $zz$
components of the shear tensor, while $\Omega$ is related to the
magnetic field. The new time $\tau$ is a dimensionless time (or
``logarithmic'' time). The quantity $H(t)$ (because of
(\ref{KMWAP2})) will have dimensions of ${\textrm{cm}}^{-1}$ and the
sign of $\tau$ becomes determined from the sign of $H(t)$. However,
we have chosen $\Omega=\beta$ and $\kappa\lambda=3H^{2}_{0}$ in
order to avoid the presence of annoying constants in the system of
equations. Inserting the equations of state (\ref{KMWAP9}) and the
new definitions (\ref{DV1}), (\ref{DV2}) into (\ref{constr1}) and
(\ref{KMWAP8}) we obtain the constraint:
\begin{equation}
(S^{y})^{2}-(S^{z})^2-S^{y}S^{z}+3\HH^{2}=3\Gamma_{U},
\label{DV3d}\end{equation}
plus the system
\begin{subequations}
\begin{eqnarray}
S^{y}_{,\tau}&=& -\beta\Gamma_{\M}-3S^{y}\HH, \label{DV3a}
\\
S^{z}_{,\tau}&=& 2\beta\Gamma_{\M}-3S^{z}\HH, \label{DV3b}
\\
\HH_{,\tau}&=&\beta \Gamma_{\M}-\frac{3}{2}(\Gamma_{p}+
\HH^{2}+\frac{(S^{y}+S^{z})^{2}}{3}-\frac{S^{y}S^{z}}{3}),
\label{DV3c}
\\
\beta_{,\tau}&=&2\beta(S^{z}-2\HH),   \label{DV3e}
\\
\mu_{,\tau}&=&\frac{1}{\Gamma_{U,\mu}}\left[(2\HH-S^{z})
(\Gamma_{\M}+2\Gamma_{U,\beta})\beta-3\HH(\Gamma_{p}+\Gamma_{U})\right]
\label{DV3f}\nonumber
\\ \
\end{eqnarray}
 \label{DV3}
\end{subequations}
Notice that, as opposed to cosmological sources and
models~\cite{TMW} where $H_{0}=0.59\times
10^{-28}{\textrm{cm}^{-1}}$ would play the role of the Hubble scale
constant, in our magnetized Fermi gas we have $H_{0}=0.86\times
10^{-12}{\textrm{cm}^{-1}}$, which is a much smaller length scale.
This is logical and consistent because it indicates that our
simplified model is examined on local scales smaller than cosmic
scales. The scale $1/H_{0}\simeq 1.15\times 10^{12}\,{\textrm{cm}}$
is the order of the distance of an astronomical unit.\\

The results mentioned in references \cite{Aurora1,Aurora2,Aurora3}
for the electron gas show that for an intense magnetic field, of the
order of the critical field $B_{c}$, all the electrons are in the
ground state of the Landau level $n=0$, and consequently we have $p_
{\perp} =0$. It is an interesting issue to study how the electron
gas evolves in this case, in which the functions $\Gamma(\beta,\mu)$
are simplified considerably, taking the following forms:
\begin{subequations}
\begin{eqnarray}
\Gamma_{p}&=&\frac{a_{_0}}{3}(\mu^2-\frac{5}{2})+\frac{1}{2}\textrm{arcsinh}(\frac{a_{_0}}{\mu})+\beta(a_{_0}-b_{_0}),\ \label{DV4a} \\
\Gamma_{\M}&=&(a_{_0}-b_{_0}), \label{DV4b} \ \ \ \ \\
\Gamma_{U}&=&a_{_0}(\mu^2-\frac{1}{2})-\frac{1}{2}\textrm{arcsinh}(\frac{a_{_0}}{\mu})+\beta(a_{_0}+b_{_0}),     \label{DV4c} \ \ \ \ \\
a_{_0}&=&{\mu\sqrt{\mu^2-1}}, \ \ b_{_0}=\ln (\mu+a_{_0}/\mu),\
c_{_0}=0, \ \alpha_{_0}=1.\nonumber \label{DV4d}
\end{eqnarray}\label{DV4}
\end{subequations}
Thus, substituting (\ref{DV4}) into (\ref{DV3}) yields a
self--consistent system of five ordinary differential equations (\ref{DV3}),
with the unknown functions $\beta$,$\HH$,$S^{y}$, $S^{z}$ and $\mu$,
and the constraint (\ref{DV3d}), which only admits a numerical solution.\\

From the equations of state (\ref{KMWAP9}) it follows that the
chemical potential must satisfy\, $\mu\geq 1$, which is typically
correct for systems with densities of the order $\sim
10^{7}{\textrm{gm/cm}}^{3}$ or larger. For white dwarfs or neutron
stars the chemical potencial takes values around $\sqrt{3}\simeq
1.732$. Since $U > 0$, then the chemical potential $\mu\geq 1$ and
from (\ref{DV3d}) we obtain the constraint equation:
\begin{equation}
-(S^{y})^{2}-(S^{z})^{2}-S^{y}S^{z}+3\HH^{2} = 3 \Gamma_{U}\geq 0.
\label{DV5}
\end{equation}
so that our physical 5-dimensional phase space is restricted by the relations:
\begin{subequations}
\begin{eqnarray}
3\HH^{2} &\geq& (S^{y})^{2}+(S^{z})^{2}+S^{y}S^{z}, \\
 \mu^2&\geq& 1+2\beta,
\\
 \beta &\geq& \frac{3\Gamma_{p}-\Gamma_{U}}{2\Gamma_{\M}}. \label{DV6}
\end{eqnarray}
\end{subequations}
For non--tilted Bianchi-I models there is only one independent
component of the shear tensor, which we are taking to be $S^{z}$,
however it is useful to use also the $S^{y}$ component for several
numerical calculations. Also, the form of the components of the
shear tensor determines the form of the metric coefficients. From
(\ref{KMWAP2})--(\ref{KMWAP3}) we get:
\begin{equation}
\frac{A_{,\tau}}{A}=(S^{x}+\HH),  \ \
\frac{B_{,\tau}}{B}=(S^{y}+\HH),  \ \
\frac{C_{,\tau}}{C}=(S^{z}+\HH).  \ \
\label{DV7}
\end{equation}
where $ S^{x}=-S^{y}-S^{z}.$

\section{Numerical Results and Discussions}

Using (\ref{KMWAP2}) and (\ref{DV1}) we can express the local average volumen
$\CV\equiv ABC$ in terms of $\HH$ and the dimensionless time $\tau$ as
\begin{equation}
\CV(\tau)=\CV(0)\,\exp\left(3\,{\int}^{\tau}_{\tau_0}{\HH d{\tau}}\right). \label{NRAD1}
\end{equation}
clearly illustrating why the time $\tau$ is known as a
``logarithmic'' time. Notice that the sign of $\HH(\tau)$ denotes
expanding ($>0$) or collapsing ($<0$) local volumes.
\begin{figure}
\centering
\includegraphics[height=8cm,width=8cm,angle=270]{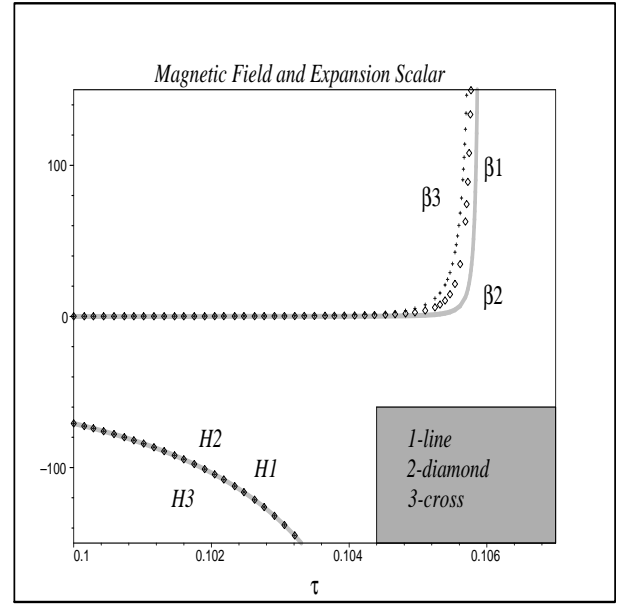}
\caption{ Numerical solutions for the magnetic field $\beta(\tau)$
and the expansion scalar $\HH(\tau)$. Here we got three different
initial conditions $S^{x}(0)=0,S^{y}(0)=-1, S^{z}(0)=1, \mu(0)=2,$
and
$\beta_{1}(0)=10^{-5},\beta_{2}(0)=5.10^{-5},\beta_{3}(0)=10^{-4}$
respectively, for the initial magnetic field. Then because of
(\ref{DV5}), the initial expansions should take the values
$\HH_{1}(0)=\HH_{2}(0)=\HH_{3}(0)=-4.82$. The numerical results for
the collapse times are $\tau_{1}=0.1059, \tau_{2}=0.1058$, and
$\tau_{3}=0.1057$ respectively. } \label{FIG0}
\end{figure}
\subsection{Singularities}

Since the equations of state that we are considering are associated
with compact objects of very high density (at least $\sim 10^{7}
\textrm{gm/cm}^3$), the evolution range of the models for diluting
lower densities is not physically interesting and will not be
pursued any further.  This means that we will only consider the
collapsing phase of the models, so that we will only need to examine
initial conditions in which the initial expansion $\HH_{0}$ is
negative. We tested numerically the models using a wide range of
different initial conditions  covering the full range of physically
interesting  values for compact objects, from white dwarf to neutron
stars, for example: $\mu_{0}=2$, corresponding to densities of $\sim
10^{7}g/cm^{3}$, while $\beta_{0}=10^{-5}$ represents magnetics
fields of $10^{8}$ G. Together with $\HH_{0}<0$, we considered in
particular: $S^{y}_{0}=0,\pm 1$ and $S^{z}_{0}=0,\pm 1$,
corresponding to cases of zero initial deformation and initial
deformation on the $y$ or $z$ directions.

As long as $\HH_{0}<0$ the models exhibit a general collapsing
behavior $\HH \rightarrow-\infty$, independent of the initial values
of other functions (see an example in figure(\ref{FIG0})). In all
collapsing configurations the magnetic field intensity diverges to
infinity regardless of its initial value. In figure(\ref{FIG0}) we
show several numeric curve solutions for different values of the
initial magnetic field. We can see how for increasing intensity
(labeled by $\beta1< \beta2< \beta3$ in the figure) the expansion
shows a faster decay to $-\infty$. Consequently, whenever we
increase the initial magnetic field the collapse times decrease.

\begin{figure}
\centering
\includegraphics[height=8cm,width=8cm,angle=270]{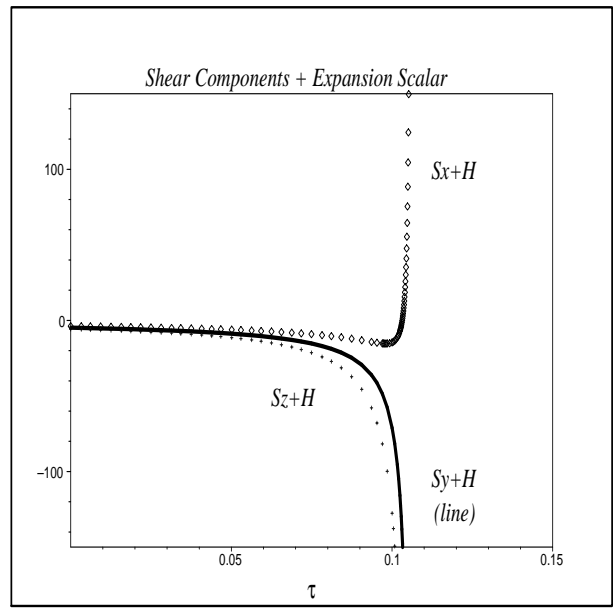}
\caption{ Trajectories of the functions $S^{i}+\HH$ for $i=x,y,z$.
We take the following initial conditions: \,
$S^{x}(0)=1,\,S^{y}(0)=0, \,S^{z}(0)=-1, \,\mu(0)=2,\,\beta(0)=0,$
and\, $\HH(0)=-3.43$. This system shows an initial deformation in
the $x$ direction with the magnetic field pointing in the $z$
direction. The collapse is in the form of a ``line'' singularity in
the $x$ direction.} \label{FIG1}
\end{figure}

We consider initial condition with initial shear deformation not
necessarily in the direction of the magnetic field, so that
anisotropic  singularities of type ``cigar'' emerge along any of the
three axes, that is, a collapse that can be parallel or
perpendicular to the magnetic field. An examination of all these
cases reveals that the collapse state strongly depends on  the
magnitude of the initial shear. For example, if we have a large
initial shear (deformation) in the $x$ direction, say: $S^{x}_{0}\gg
S^{y}_{0},S^{z}_{0},\beta_{0},\mu_{0}$, then a type cigar
singularity emerges along the $x$ direction. This is shown in
figure(\ref{FIG1}), illustrating (by means of the set of equations
(\ref{DV7})) that the metric function $A$ tends to infinity, while
the other metric functions, $B$ and $C$ rapidly fall to zero. In
general, the initial configuration of the system and the initial
values of the shear tensor determinate the privileged direction of
the anisotropic collapse.

The numerical trials also show that there is always a threshold
value for the initial magnetic field intensity that influences the
direction of the ``cigar'' type singularity. This is illustrated in
figure(\ref{FIG1}): if we increase the initial magnetic field
intensity to $\beta_{0}=1$ then we will obtain a ``cigar'' type
singularity along the $z$ direction for any $\beta_{0}\geq 1$.
However, $\beta_{0}=1 \sim 10^{13}$ G, which is not a physically
realistic value in magnetic white dwarfs but could be reasonable in
a primordial magnetized universe model.

An isotropic ``point'' singularity can always emerge for
configurations with zero initial deformation, {\it i.e.} \,
$S^{x}_{0}=S^{y}_{0}=S^{z}_{0}=0$ and $\beta_{0}=0$. However, even
with zero initial deformation there is always a threshold value of
initial magnetic field intensity for which the singularity becomes
anisotropic along the $z$ direction.
\subsection{Phase Space and Critical Subspaces}
As mentioned previously, the Einstein--Maxwell system can be written
as an autonomous system associated with a 4-dimensional phase space
in the variables $(S^{z},\beta,\mu,\HH)$. Notice that $\Sigma^{y}$
can always be found if we determine $\Sigma^{z}=\Sigma$, which is
the only independent shear component. Considering the four
primordial functions the system will take the form:
\begin{subequations}
\begin{eqnarray}
\dot{U}&=&-(U+p-\frac{2}{3}\B\M)\Theta-\B\M\Sigma, \\
\dot{\Sigma}&=&\frac{2}{3}\kappa\B\M-\Theta\Sigma,\\
\dot{\Theta}&=&\kappa(\B\M+\frac{3}{2}(U-p))-\Theta^2,\\
\dot{\beta}&=&\frac{2}{3}\beta(3\Sigma-2\Theta).
\end{eqnarray}
\end{subequations}
where $\Sigma=\sigma^{z}$. If we work with the dimensionless functions defined
in (\ref{DV2}) this system becomes:
\begin{subequations}
\begin{eqnarray}
S^{z}_{,\tau}&=& 2\beta \Gamma_{\M}-3\HH S^{z},\\
\HH_{,\tau}&=& \beta\Gamma_{\M}+\frac{3}{2}(\Gamma_{U}-\Gamma_{p})-3\HH^{2}, \label{PSaCS3b}\\
\beta_{,\tau}&=& 2\beta(S^{z}-2\HH),\\
\mu_{,\tau}&=&\frac{1}{\Gamma_{U,\mu}}[(2\HH-
S^{z})(\Gamma_{\M}+2\Gamma_{U,\beta})\beta-3\HH(\Gamma_{p}+\Gamma_{U})].
\nonumber \\
\end{eqnarray}\label{PSaCS3}
\end{subequations}
Note that only change the equation for $\HH(\tau)$, but we can
reduce the equations (\ref{PSaCS3b}) or (\ref{DV3c}) to the
constraint (\ref{DV3d}), therefore both systems are equivalent.
Then, (\ref{DV3a}) is only necessary for the computation of the
metric coefficients.

\begin{figure}
\centering
\includegraphics[height=8cm,width=8cm,angle=270]{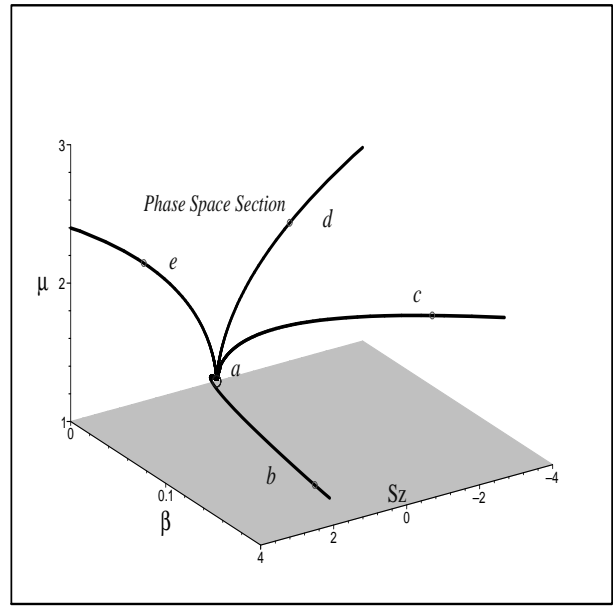}
\caption{ Trajectories in a section of the phase space
($S^{z},\beta,\mu$) for four different initial conditions. The
shaded $\mu=1$ plane is bounded by $-4<S^{z}<4$ and $0<\beta<0.2$.
The stable attractor is the point marked by $a$, and the curves
$b,c,d$ and $e$ are numerical solutions on the 3-dim section of the
phase space. All the trajectories here fall from $\tau=0$ toward the
stable attractor $a$. Similarly, from $\tau=0$ (empty spheres) all
them scape toward the singularity.} \label{FIG2}
\end{figure}

In figure(\ref{FIG2}) we represent a 3-dimensional section of the
phase space $(S^{z},\beta,\mu)$ with different curves for several
initial conditions. As shown in the figure, the initial value of the
expansion $H_{0}$ determines the global evolution of the numerical
solutions curves. All curves  starting at $\tau=0$ with initial
expansion $H_{0}=-\sqrt{\kappa\lambda/3}$ converge into the stable
attractor marked as ``$a$'', whereas if we choose
$H_{0}=\sqrt{\kappa\lambda/3}$ and start at $\tau=0$, the curves
evolve towards the anisotropic singularity. Setting the left hand
sides of the equations in (\ref{PSaCS3}) to zero and solving the
algebraic system we find the set of critical points associated with
this system, including the stable attractor marked as ``$a$''. These
points are given by:
\begin{equation}
a=\{S^{z} = 0, \beta = 0, \mu = 1, \HH = 0\}, \label{CS1a}
\end{equation}\label{CS1}
For the 4-dim phase space $(S^{z},\beta,\mu,\HH)$ we have 4 possible
3-dim sections of the same phase space. We have computed this
numerical solutions with similar results.
\section{Conclusion.}

We have presented a model based on the dynamical description of a
local volume of a magnetized, self-gravitating, Fermi gas in the
basic Landau level $n=0$. Since we are considering complicated
equations of the state, we have worked with the simplified form of
the Einstein-Maxwell equations that follow by assuming a Bianchi-I
space time represented by a Kasner metric, whose source of
anisotropy is just the magnetic field. This simplified spacetime
geometry provides a convenient toy model for a rough understanding
the local collapsing behavior of the type of matter found inside a
magnetized star like a white dwarf or a neutron star.

The relevance of the present paper emerges from our study of the
collapsing singularities, which can be isotropic point--like or
anisotropic of type cigar. Point singularities emerge under very
special initial conditions of zero magnetic field, zero shear
deformation or both. Cigar type singularities can also be obtained
in all directions, depending on the initial values of the shear
deformation. However, for an initial magnetic field intensity having
a sufficiently large value the end singularity always becomes of
type cigar in the direction of the field.
 This result is important because the value of the magnetic field determines
the type of collapse and this is in agreement with the
non-relativistic previous paper which examined the collapse of this
type of magnetized gases within a Newtonian framework
\cite{Aurora1},\cite{Aurora2},\cite{Aurora3}.

 As discussed in \cite{SBC} by Collins $\&$ Ellis, orthogonal Bianchi
models like the one we are considering are globally hyperbolic and
only present a single singularity, so that hypersurfaces of constant
time (orthogonal to the 4- velocity) are global Cauchy hypersurfaces
and every point in spacetime can be causally connected to the
latter. The 4-velocity is a geodesic field and the singularity is
marked by a specific constant time value, so that it is non-timelike
and every event in spacetime can be causally connected to this
singularity (in particular by the timelike geodesics that are
integral curves of the 4-velocity field).

Now, in this article we are only considering a collapsing regime
from an initial hypersurface of constant time. Thus, every future
directed timelike curve (geodesic or not) starting at any initial
Cauchy hypersurface of constant time will terminate in the collapse
singularity. Under these conditions, this singularity is obviously
censored.

\acknowledgements We gratefully acknowledge very helpful through the
financial support from grant PAPIIT--DGAPA--IN117803. The authors
acknowledge the Office of External Activities of ICTP for its
support through NET-35. The authors are also indebted to Dr. Hugo
P\'{e}rez Rojas by the permanent interest in this work.

\end{document}